# Ti SURFACE MODIFICATION FOR BIOMEDICAL APPLICATIONS

Alexander Sobolev[1], Michael Zinigrad[1], Konstantin Borodianskiy[1,*]

[1] *Zimin Advanced Materials Laboratory, Department of Chemical Engineering, Biotechnology and Materials, Ariel University, Ariel 40700, Israel*

\* Correspondence: konstantinb@ariel.ac.il; Tel.: +972-3-9143085

**ABSTRACT**

Micro Arc Oxidation (MAO) is an electrochemical approach for the surface treatment usually applied on so called valve metals such as Al, Mg and Ti. MAO is usually carried out an aqueous electrolyte, which involves a bath cooling and leads to the creation of surface contained components originated in the electrolyte. In current work, we applied a different approach of ceramic surface formation on Ti alloy used in biomedical applications. Here, MAO process conducted in molten nitrate salt at 280 °C. The developed surface morphology, chemical and phase composition, and corrosion resistance were investigated and described in the work.

**Keywords:** micro arc oxidation; Ti alloy; molten salt; biomedical application.

## 1. INTRODUCTION

MAO is traditionally used to produce multipurpose wear-, corrosion- and heat-resistant, dielectric and decorative coatings on the valve metals, such as Al, Mg, Ti, Ta, Nb, Zr and Be [1 - 5].

Titanium is the attractive metal due to its high specific strength [6], corrosion resistance [7] and excellent biocompatibility [8]. Some research works have showed Ti alloys coating formation by MAO for tribological, biomedical, dielectric, and photovoltaic coatings [9-13]. One of the most applicable alloys for biomedical applications is Ti-6Al-4V which contains stabilizers for both α and β phases [14-16].

MAO is basically based on the anodizing electrochemical reaction, which takes place on a metallic surface accompanied by microarc discharges which form oxide ceramic surface layer with particular morphology and phase composition [17]. This is relatively simple and environmentally friendly technology.

However, some disadvantages are found in MAO process in aqueous electrolyte such as the temperature of the electrolyte, the duration of the treatment process, chemical composition and structure of the substrate, and the electrical parameters of the process [18, 19]. In our previous works we have shown application of molten salt electrolyte as preferable type of electrolyte in MAO process [20-22]. We found that the obtained surfaces on Al and Ti alloys contain oxide phases only and the process is energy preferable compared to the traditional MAO in aqueous solutions.

In present work, we examined formation of oxide coating in molten nitrate salts by MAO process on Ti alloys for biomedical applications. Chemical and the phase composition of the coating as well as its morphology and corrosion resistance were investigated and shown in the work.

## 2. EXPERIMENTAL

Ti-6Al-4V alloy specimens (chemical composition shown in Table 1) were grounded using abrasive papers grits up to #4000, and then subjected to the ultrasonic cleaning in acetone.

**Table 1.** Chemical composition of the alloy Ti-6Al-4V.

| Chemical Element, % mass | | | |
|---|---|---|---|
| V | Fe | Al | Ti |
| 4 | 0.11 | 6 | Bas |

MAO treatment was performed at 280 °C in the eutectic $KNO_3$ - $NaNO_3$ (Sigma-Aldrich) electrolyte. The process was conducted in a Ni crucible which also served as a counter electrode. The surface ratio of anode-to-cathode was 1:30, the anodic current density was 250mA/cm$^2$, and the voltage was limited by the galvanostatic mode. The applied power supply parameters were as follow: $I_{max}$= 35A, $U_{max}$=1000V, current and voltage were pulsed with a square-wave sweep at a frequency of 1Hz ($t_a$=$t_k$=0.5s) by a pulse generator Digit-EL PG-872. The duration of the treatment was 10min at a growth rate of 0.25µm/min. Behavior of current vs. time and voltage vs. time were recorded by Fluke Scope Meter 199C (200 MHz, 2.5 GS s$^{-1}$).

Microstructural examinations of the obtained coatings were done on the cross-sections of the treated and untreated specimens by TESCAN MAIA3 electron microscope equipped with an energy dispersive X-ray spectroscopy (EDS) system by Oxford instruments with an X-Max$^N$ detector. The phase analysis was determined by the X'Pert Pro diffractometer (PANalytical B.V.) with Cu$_\alpha$ radiation (λ=1.542Å) at the grazing incidence mode (angle of 3°) at 40kV and 40mA.

The corrosion behavior of the obtained samples was examined by potentiodynamic polarization test in a 3.5wt% NaCl (Sigma-Aldrich Co.) solution by PARSTAT 4000A potentiostat/galvanostat. Three- electrode cell configuration was used for the corrosion test where a Pt sheet acted as a counter electrode and saturated Ag/AgCl (Metrohm Autolab B.V.) acted as a reference electrode. The polarization resistance of a sample was determined at the range of ±250mV with the respect to the recorded corrosion potential at a scan rate of 0.1mV/s. Prior to the potentiodynamic polarization test, specimens were held in the 3.5wt% NaCl solution for 1h.

## 3. RESULTS AND DISCUSSION

### 3.1. Morphology and elemental analysis

The surface morphology and the chemical compositions of the MAO treated specimens were investigated by electron microscopy and EDS, respectively. The corresponding images are shown in Figure 1 (a, b).

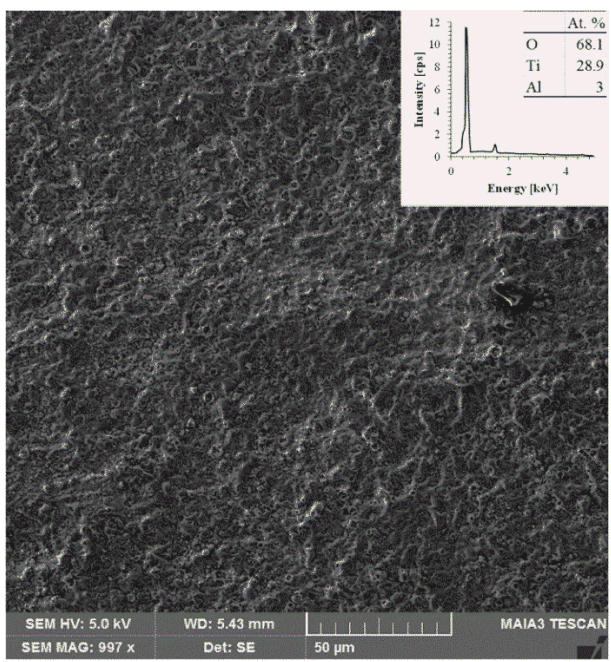

**Figure 1.** Electron microscopy image of the alloy Ti-6Al-4V surface obtained after MAO treatment and its elemental composition obtained by EDS.

Surface morphology observation revealed that the developed coating has no cracks indicating a low cooling rate of the coating formation. The obtained surface consists of round-shaped pores with the diameter ranging from 0.15μm up to 0.5μm. Elemental analysis detected appearance of the $TiO_2$ composition and traces of Al which are originated from the alloy. No additional elements were detected.

*3.2. Phase analysis*

XRD pattern of Ti alloy surface treated by MAO process is shown in Figure 2.

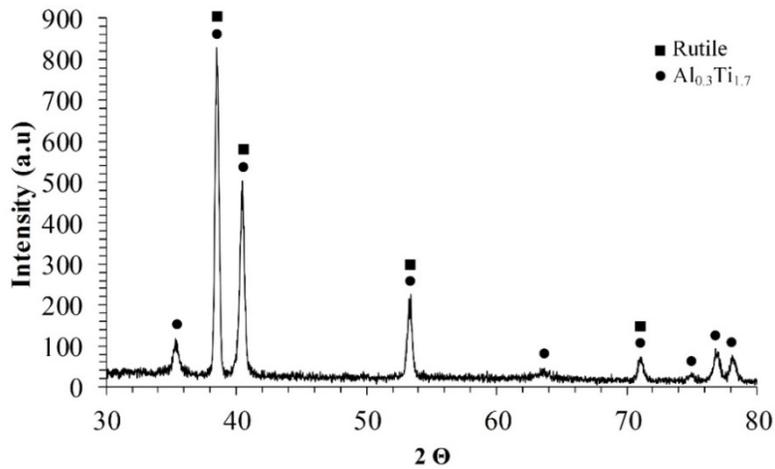

**Figure 2.** XRD pattern of the obtained surfaces after MAO treatment.

XRD investigations evaluated rutile and intermetallic of $Al_{0.3}Ti_{1.7}$ phases on the surface of the treated alloy. Rutile was formed as the result of the oxidation process and Ti/Al intermetallic phase was formed resulting in Al noticeable presence in the alloy subjected to the treatment. Additionally, no new phases were formed during the process.

Microstructure of cross section and the line scan elemental analysis of the Ti alloy treated by MAO process are shown in Figure 3.

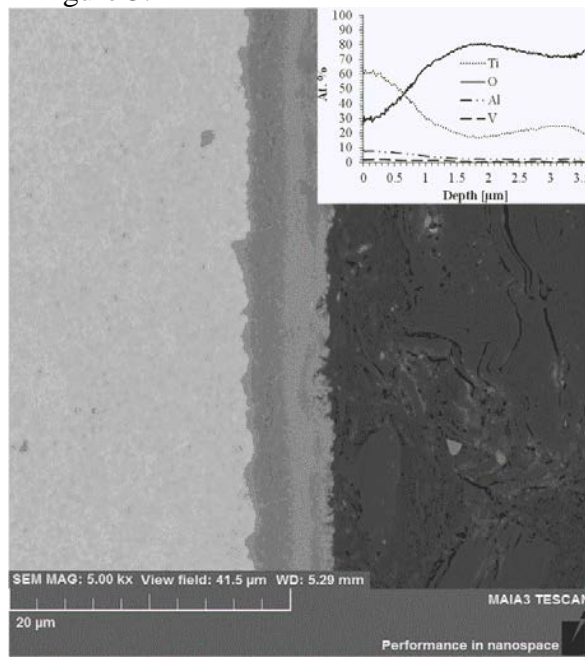

**Figure 3.** Electron microscopy image of cross section of Ti alloy after MAO treatment and its line scan elemental analysis.

Electron microscopy image jointly with the line scan investigations evaluated that the obtained oxide coating is uniformed with a thickness of 2.5 µm. The elemental analysis detected elements which fit the composition of the expected oxide coating.

*3.3. Corrosion resistance investigation*

The corrosion properties of the treated specimens were determined by the potentiodynamic polarization method. The obtained curves are presented in Figure 4.

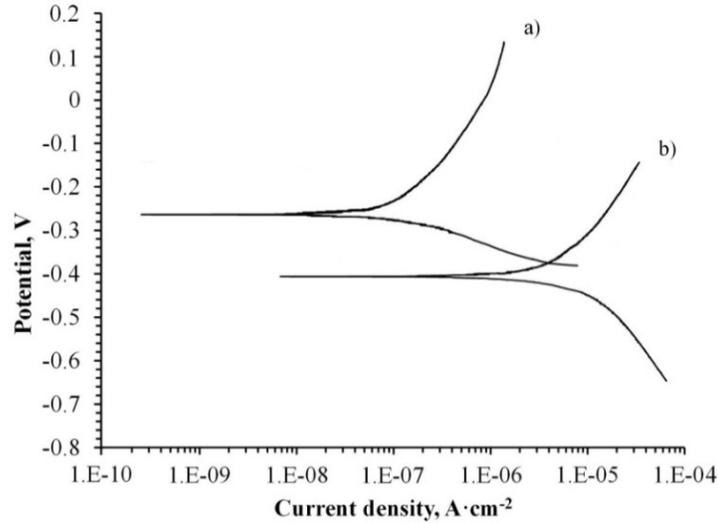

**Figure 4.** Potentiodynamic polarization curves for alloy Ti alloy after MAO treatment (a) and untreated alloy (b).

The obtained curves of corrosion resistance behavior are presented as semi-logarithmic coordinates. It is evident that a higher corrosion resistance is obtained once the corrosion potential is higher and corrosion current density is lower. Therefore, a corrosion potential of the coated specimen was shifted to a more positive and the current density to a more negative values. Probably it may indicate the reduction of the anodic and the cathodic processes due to the presence of a new developed oxide coating. Movement of the corrosion potential towards the anodic area shows a resistance improvement of the treated specimen.

The polarization resistance ($R_p$) was calculated according to equation 1:

$$R_p = \frac{\beta_a \times \beta_c}{2.3 \times i_{corr}(\beta_a+\beta_c)} \qquad (1)$$

where $\beta_a$ and $\beta_c$ were calculated from slopes of the anodic and the cathodic curves on the plot in Figure 4. Calculation results are also summarized in Table 2.

**Table 2.** Calculated corrosion test results of Ti alloy after MAO treatment and untreated alloy.

| Samples | $E_{corr}$ [mV] | $i_{corr} \times 10^{-6}$ [A] | $\beta_a$ [mV/decade] | $\beta_c$ [mV/decade] | $R_p \times 10^3$ [$\Omega/cm^2$] |
|---|---|---|---|---|---|
| Untreated alloy Ti-6Al-4V | -398 | 6.95 | 395 | 316 | 10.98 |
| Treated alloy Ti-6Al-4V | -260 | 0.16 | 390 | 97 | 213.76 |

Calculations results presented in Table 2 demonstrate that the polarization resistance of the treated specimen was 213.76 kΩ/cm$^2$ while the untreated specimen reached value of 10.98 kΩ/cm$^2$. These values exhibit that oxide coating on Ti alloy is almost 20 times higher than the untreated alloy.

## 4. CONCLUSIONS

In the presented work a new approach of obtaining ceramic coating on Ti alloy for biomedical applications was shown. The coating was obtained by MAO process carried out in electrolyte based on the eutectic composition of the nitrate salts.

Morphological investigations evaluated formation of 2.5µm oxide coating contained homogeneously distributed round-shaped pores. XRD analysis evaluated presence of titanium oxide and titanium aluminum intermetallic ($Al_{0.3}Ti_{1.7}$) phases. These results were also confirmed by EDS analysis. The corrosion test revealed that resistance of the coated alloy 20 times higher than the untreated one.

**Acknowledgments:** This work was carried out with the support of the Ministry of Aliyah and Integration, the State of Israel.

### REFERENCES

1. Yerokhin A.L., Nie X., Leyland A., Matthews A.: Characterization of oxide films produced by plasma electrolytic oxidation of a Ti–6Al–4V alloy. *Surf. Coat. Technol.* **2000**, 130, 195–206.
2. Wang Y., Jiang B., Lei T., Guo L.: Dependence of growth features of microarc oxidation coatings of titanium alloy on control modes of alternate pulse. *Mater. Lett.* **2004**, 58, 1907–1911.
3. Fei C., Hai Z., Chen C., Yangjian X.: Study on the tribological performance of ceramic coatings on titanium alloy surfaces obtained through microarc oxidation. *Prog. Org. Coat.* **2009**, 64, 264–267.
4. Huang P., Wang F., Xu K., Han Y.: Mechanical properties of titania prepared by plasma electrolytic oxidation at different voltages. *Surf. Coat. Technol.* **2007**, 201, 5168–5171.
5. Santos-Coquillat A., Gonzalez Tenorio R.: Mohedano M., Martinez-Campos E., Arrabal R., Matykina E., Tailoring of antibacterial and osteogenic properties of Ti6Al4V by plasma electrolytic oxidation. *Appl. Surf. Sci.* **2018**, 454, 157–172.
6. Niinomi M.: Mechanical properties of biomedical titanium alloys. *Mater. Sci. Eng., A* **1998,** 243, 231–236.
7. Gu Y., Ma A., Jiang J., Li H., Song D., Wu H., Yuan Y.: Simultaneously improving mechanical properties and corrosion resistance of pure Ti by continuous ECAP plus short-duration annealing. *Mater. Charact.* **2018**, 138, 38–47.

8. Geetha M., Singh A.K., Asokamani R., Gogia A.K.: Ti based biomaterials, the ultimate choice for orthopaedic implants – A review. *Prog. Mater. Sci.* **2009**, 54, 397–425.
9. Lederer S., Lutz P., Fürbeth W.: Surface modification of Ti 13Nb 13Zr by plasma electrolytic oxidation. *Surf. Coat. Technol.* **2018**, 335, 62–71.
10. Huang P., Wang F., Xu K., Han Y.: Mechanical properties of titania prepared by plasma electrolytic oxidation at different voltages. *Surf. Coat. Technol.* **2007**, 201, 5168–5171.
11. Wheeler J.M., Collier C.A., Paillard J.M., Curran J.A.: Evaluation of micromechanical behaviour of plasma electrolytic oxidation (PEO) coatings on Ti–6Al–4V. *Surf. Coat. Technol.* **2010**, 204, 3399–3409.
12. Khorasanian M., Dehghan A., Shariat M.H., Bahrololoom M.E., Javadpour S.: Microstructure and wear resistance of oxide coatings on Ti–6Al–4V produced by plasma electrolytic oxidation in an inexpensive electrolyte. *Surf. Coat. Technol.* **2011**, 206, 1495–1502.
13. Fakhr Nabavi H., Aliofkhazraei M., Sabour Rouhaghdam A.: Electrical characteristics and discharge properties of hybrid plasma electrolytic oxidation on titanium. *J. Alloys Compd.* **2017**, 728, 464-475.
14. Geetha M., Singh A.K., Asokaman R., Gogia A.K.: Ti based biomaterials, the ultimate choice for orthopaedic implants – A review. *Progress in Materials Science* **2009**, 54, 397–425.
15. Yeung W.K., Reilly G.C., Matthews A., Yerokhin A.: In vitro biological response of plasma electrolytically oxidized and plasma-sprayed hydroxyapatite coatings on Ti–6Al–4V alloy. *J. Biomed Mater Res B Appl Biomater.* **2013**, 101(6), 939-949.
16. Xia L., Xie Y., Fang B., Wang X., Lin K.: In situ modulation of crystallinity and nano-structures to enhance the stability and osseointegration of hydroxyapatite coatings on Ti-6Al-4V implants. *Chem Eng J* **2018**, 347, 711-720.
17. Yerokhin A.L., Nie X., Leyland A., Matthews A., Dowey S.J.: Plasma electrolysis for surface engineering. *Surf. Coat. Technol.* **1999**, 122, 73–93.
18. Kossenko A., Zinigrad M.: A universal electrolyte for the plasma electrolytic oxidation of aluminum and magnesium alloys. *Mater. Des.* **2015**, 88, 302–309.
19. Habazakia H., Tsunekawa S., Tsuji E., Nakayama T.: Formation and characterization of wear-resistant PEO coatings formed on β-titanium alloy at different electrolyte temperatures. *Appl. Surf. Sci.* **2012**, 259, 711–718.
20. Sobolev A., Kossenko A., Zinigrad M., Borodianskiy K.: Comparison of plasma electrolytic oxidation coatings on Al alloy created in aqueous solution and molten salt electrolytes. *Surf. Coat. Technol.* **2018**, 344, 590–595.
21. Sobolev A., Wolicki I., Kossenko A., Zinigrad M., Borodianskiy K.: Coating Formation on Ti-6Al-4V Alloy by Micro Arc Oxidation in Molten Salt. *Materials* **2018**, 11 (9), 1611.
22. Sobolev A., Kossenko A., Zinigrad M., Borodianskiy K.: An investigation of oxide coating synthesized on an aluminum alloy by plasma electrolytic oxidation in molten salt. *Appl. Sci.* **2017**, 7, 889.